\documentclass[12pt]{article}

\usepackage{amssymb}
\usepackage{fullpage}
\usepackage{graphicx}
\usepackage{url}

\newtheorem{definition}{Definition}
\newtheorem{theorem}{Theorem}
\newtheorem{lemma}{Lemma}
\newtheorem{observation}{Observation}

\newcommand{\IR}{\mathbb{R}}
\newcommand{\C}{\mathcal{C}}
\newcommand{\eps}{\varepsilon}
\newcommand{\diam}{\mathord{\it diam}}

\newcommand{\qed}{\rule{0.5em}{1.5ex}}
\newcommand{\fqed}{{\hfill~\qed}}
\newenvironment{proof}{{\noindent \bf Proof.}}
                      {{\hfill \fqed} \vspace{1em}}

\title{Metric and Geometric Spanners that are Resilient to 
Degree-Bounded Edge Faults} 
\author{
Ahmad Biniaz\thanks{School of Computer Science, University of Windsor,
                Canada. Supported by NSERC.} 
\and 
Jean-Lou De Carufel\thanks{School of Electrical Engineering and Computer 
    Science, University of Ottawa, Canada. Supported by NSERC.} 
\and
Anil Maheshwari\thanks{School of Computer Science, 
                    Carleton University, Ottawa, Canada. Supported by 
                    NSERC.} 
\and 
Michiel Smid\thanks{School of Computer Science, 
                    Carleton University, Ottawa, Canada. 
                    Supported by NSERC.}} 
\date{\today}

\begin{document} 

\maketitle 

\begin{abstract} 
Let $H$ be an edge-weighted graph, and let $G$ be a subgraph of $H$. 
We say that $G$ is an $f$-fault-tolerant $t$-spanner for $H$, if the 
following is true for any subset $F$ of at most $f$ edges of $G$: 
For any two vertices $p$ and $q$, the shortest-path distance between 
$p$ and $q$ in the graph $G \setminus F$ is at most $t$ times 
the shortest-path distance between $p$ and $q$ in the graph 
$H \setminus F$. The value of $t$ is called the stretch factor. 
Numerous results are known for constructing such 
spanners, both for general graphs $H$ and for complete graphs $H$ 
whose vertex set is a set of points in $\IR^d$ or in a metric space 
of bounded doubling dimension. 

Bodwin, Haeupler, and Parter (SODA 2024) generalized this notion to the 
case when $F$ can be any set of edges in $G$, as long as the maximum 
degree of $F$ is at most $f$. For example, if $f=1$, then $F$ can be a 
perfect matching in $G$. They gave constructions for general graphs $H$. 

We present three new results for constructing spanners that are resilient 
to edge faults of maximum degree $f$. 
 
We first consider the case when $H$ is a complete graph whose vertex set 
is an arbitrary metric space. We show that if this metric space 
contains a $t$-spanner with $m$ edges, then it also contains a
graph $G$ with at most $(4f-1)m$ edges, that is resilient to edge faults 
of maximum degree $f$ and has stretch factor $O(ft)$. We also give 
an example for which our construction has stretch factor 
$\Omega(ft)$. If $f=1$, the graph $G$ has at most $3m$ edges and 
its stretch factor is at most $3t$. For this case, we give an example 
for which $G$ has stretch factor arbitrarily close to $3t$. 

Next, we consider the case when $H$ is a complete graph whose vertex 
set is a metric space that admits a well-separated pair decomposition. 
We show that, if the metric space has such a decomposition of size 
$m$, then it contains a graph with at most $(2f+1)^2 m$ edges, 
that is resilient to edge faults of maximum degree $f$ and has stretch 
factor at most $1+\eps$, for any given $\eps > 0$. For example, if
the vertex set is a set of $n$ points in $\IR^d$ ($d$ being a 
constant) or a set of $n$ points in a metric space of bounded doubling 
dimension, then the spanner has $O(f^2 n)$ edges. 

Finally, for the case when $H$ is a complete graph on $n$ points in 
$\IR^d$, we show how natural variants of the Yao- and $\Theta$-graphs 
lead to graphs with $O(fn)$ edges, that are resilient to edge faults of 
maximum degree $f$ and have stretch factor at most $1+\eps$, for any 
given $\eps > 0$. 
\end{abstract} 

\section{Introduction}   
Let $H$ be an undirected ``host graph'', in which each edge has a weight. 
For a real number $t \geq 1$, a subgraph $G$ of $H$ is called a 
$t$-\emph{spanner} for $H$, if for any two vertices $p$ and $q$ of $H$, 
\[ \delta_G(p,q) \leq t \cdot \delta_H(p,q) , 
\]
where, for any edge-weighted graph $X$, $\delta_X(p,q)$ denotes the 
length of a shortest path between $p$ and $q$ in $X$. 
The smallest such value for $t$ is called the \emph{stretch factor} 
of $G$. 

There is a large amount of literature on the problem of designing 
algorithms that compute a $t$-spanner for a graph $H$, where the goal 
is to obtain the best trade-offs between the stretch factor $t$ and 
the number of edges in $G$. For general edge-weighted graphs $H$, 
we refer the reader to the survey paper 
Ahmed \emph{et al.}~\cite{abd-gs-20}. 

In the \emph{metric} case, the host graph $H$ is a complete graph $K_S$ 
induced by a finite metric space $(S,| \cdot |)$: The vertex set of 
$K_S$ is $S$ and for any two distinct points $p$ and $q$ in $S$, $K_S$ 
contains the edge $\{p,q\}$ with weight $|pq|$. In this case, a 
graph $G$ with vertex set $S$ is a $t$-spanner for $S$, if for any 
two points $p$ and $q$ of $S$, 
\[ \delta_G(p,q) \leq t \cdot |pq| . 
\]
For several classes of metric spaces, constructions are known that 
compute, for any constant $t>1$, a $t$-spanner with $O(n)$ edges, 
where $n$ is the number of points in $S$. For example, this is the 
case for Euclidean spaces, where $S$ is a finite set of points in 
$\IR^d$ (where the dimension $d$ is a constant) and the distance 
between two points is their Euclidean distance. This was first shown 
by Keil~\cite{k-aceg-88}. Many other constructions can be found 
in Narasimhan and Smid~\cite{ns-gsn-07}. 
Another example is the case when the metric space $(S,| \cdot |)$ has 
bounded doubling dimension. For this case, 
Har-Peled and Mendel~\cite{hm-fcnld-05} were the first to show how to 
compute a $t$-spanner with $O(n)$ edges. Other results can be found 
in Filtser and Solomon~\cite{fs-gs-20}.  

\subsection{Fault-Tolerant Spanners} 
Spanners were generalized by 
Levcopoulos \emph{et al.}~\cite{lns-eacft-98,lns-iacft-99} to 
\emph{fault-tolerant} spanners. Let $f \geq 1$ be an integer. A 
subgraph $G$ of the host graph $H$ is an $f$-fault-tolerant $t$-spanner 
for $H$, if the following holds for any subset $F$ of at most $f$ 
edges of $G$: For any two vertices $p$ and $q$ of $H$, 
\begin{equation} 
\label{eqf} 
   \delta_{G \setminus F}(p,q) \leq 
           t \cdot \delta_{H \setminus F}(p,q) , 
\end{equation} 
where, for any edge-weighted graph $X=(V_X,E_X)$, $X \setminus F$ 
denotes the graph with vertex set $V_X$ and edge set $E_X \setminus F$. 
Thus, $\delta_{G \setminus F}(p,q)$ denotes the length of a shortest 
path between $p$ and $q$ in the graph obtained by removing all edges 
in $F$ from $G$. We remark that there are also versions of 
fault-tolerant spanners in which up to $f$ vertices and edges can be 
removed.  

Again, there is a large amount of literature on the problem of
constructing fault-tolerant spanners. For the case of general 
edge-weighted graphs, the first results were given by 
Chechik \emph{et al.}~\cite{clpr-fts-10}. See 
Bodwin \emph{et al.}~\cite{bdr-eft-22} for more recent results.  
For Euclidean spaces and metric spaces of bounded doubling dimension, 
we refer the reader to 
\cite{clnsr-ds-15,hp-rs-23,l-nrftg-99,l-nrgst-99,ns-gsn-07,s-ofts-14}.

Bodwin, Haeupler, and Parter~\cite{bhp-23} introduced a 
more general notion of fault-tolerant spanners. A subgraph $G=(V_H,E)$ 
of the host graph $H=(V_H,E_H)$ is an $f$-\emph{faulty-degree} 
$t$-\emph{spanner} for $H$, if for any subset $F$ of $E$, for which the 
maximum degree of the graph $(V_H,F)$ is at most $f$, and for any two 
vertices $p$ and $q$ in $V_H$, (\ref{eqf}) holds. 

For example, in a $1$-faulty-degree $t$-spanner, we are allowed to 
remove a perfect matching that has $|V_H|/2$ edges, whereas in a 
$1$-fault-tolerant $t$-spanner, we are only allowed to remove one edge. 
This shows that an $f$-faulty-degree spanner is a much stronger 
condition.

Bodwin, Haeupler, and Parter~\cite{bhp-23} proved the following 
result. Let $H$ be any edge-weighted graph with $n$ vertices. For any 
positive integers $f$ and $k$, there exists an $f$-faulty-degree 
$(2k-1)$-spanner for $H$ that has 
$O( k^k \cdot f^{1-1/k} \cdot n^{1+1/k} )$ edges. 
Notice that there is a large dependance on $k$. In particular, the 
upper bound on the number of edges is subquadratic in $n$ only if
$k = o(\log n / \log\log n)$.\footnote{Recall that $k^k = n$ solves to 
$k \sim \log n / \log\log n$.}

\subsection{Our Results} 
We consider faulty-degree spanners for the case when the host graph 
is the complete graph $K_S$ induced by a finite metric space 
$(S,| \cdot |)$. For completeness, we state the definition of a 
faulty-degree spanner for this case. 

\begin{definition}
{\em   
Let $(S,| \cdot |)$ be a finite metric space, 
let $f \geq 0$ be an integer, let $t \geq 1$ be a real
number, and let $G=(S,E)$ be an undirected graph, in which each 
edge $\{p,q\}$ has weight $|pq|$. The graph $G$ is called an 
$f$-\emph{faulty-degree} $t$-\emph{spanner} for $S$,  
if the following holds for any 
subset $F$ of $E$, for which the maximum degree of the graph 
$(S,F)$ is at most $f$: For any two points $p$ and $q$ in $S$, 
\[ \delta_{G \setminus F} (p,q) \leq t \cdot 
          \delta_{K_s \setminus F}(p,q) . 
\]
}
\end{definition} 

In Section~\ref{secGMS}, we consider arbitrary metric 
spaces $(S,| \cdot |)$. We will show the following: Assume that 
$G'=(S,E')$ is an arbitrary $t$-spanner for $S$. Let $n$ denote the 
number of points in $S$, and let $m$ denote the number of edges in $E'$. 
For any integer $f \geq 1$, we can convert $G'$ to a graph $G=(S,E)$ 
with at most $(4f-1)m$ edges, such that $G$ is an $f$-faulty-degree 
$t'$-spanner for $S$, where $t'$ is a function of $f$ and $t$. 
The construction is valid for any $f$ with $1 \leq f \leq (n-1)/2$. 
In Section~\ref{secFM}, we consider the case when $f=1$ and prove that 
$t' \leq 3t$. We also give an example for which $t'$ is arbitrarily close 
to $3t$. In Section~\ref{secgen}, we consider the general case  
$f \geq 1$ and prove that $t' \leq (8f+2)t$. For $f \geq 3$, we 
give an example for which $t' = \Omega(ft)$. 

In Section~\ref{secWSPD}, we assume that $S$ is a set of $n$ points in 
a metric space that admits a well-separated pair decomposition 
(WSPD). Let $m$ denote the size of such a WSPD. We will show that, for 
any integer $f \geq 1$ and any real constant $\eps>0$, there exists 
an $f$-faulty-degree $(1+\eps)$-spanner for $S$ that has at most 
$(2f+1)^2 m$ edges. We remark that our spanner is obtained by modifying 
the construction in Levcopoulos \emph{et al.}~\cite{lns-iacft-99} 
for standard fault-tolerant spanners. Examples of metric spaces for which 
$m=O(n)$ are Euclidean spaces of constant dimension (see Callahan and 
Kosaraju~\cite{ck-dmpsa-95}) and metric spaces of bounded doubling 
dimension (see Har-Peled and Mendel~\cite{hm-fcnld-05}). 

In Section~\ref{secEMS}, we will show that in Euclidean space, we can do 
even better: Assume that $S$ is a set of $n$ points in $\IR^d$, where 
the dimension $d$ is a constant, and the distance $|pq|$ between two 
points $p$ and $q$ is their Euclidean distance. For this case, we will 
define two graphs with vertex set $S$ that are variants of the 
Yao-graph (see Flinchbaugh and Jones~\cite{fj-scdnng-81} and 
Yao~\cite{y-cmstk-82}) and the $\Theta$-graph (see 
Clarkson~\cite{c-aaspm-87} and Keil~\cite{k-aceg-88}) and that depend 
on the maximum degree bound $f$ of any set $F$ of faulty edges.  
We will prove that, for any real constant $\eps>0$, each of these 
graphs is an $f$-faulty-degree $(1+\eps)$-spanner for $S$ with $O(fn)$ 
edges. In particular, the stretch factor does not depend on $f$. 
Observe that this is optimal, up to constant factors, because 
every vertex in an $f$-faulty-degree spanner must have degree at 
least $f$. We remark that our graphs are variations of the standard 
fault-tolerant spanners obtained by Lukovszki~\cite{l-nrftg-99}.

We conclude this section by introducing notation that will be used 
throughout the paper. 
Let $P=(p_1,p_2,\ldots,p_k)$ be a sequence of points in a finite 
metric space $(S,| \cdot |)$. The \emph{length} of this sequence is 
denoted by $|P|$ and is equal to $\sum_{i=1}^{k-1} | p_i p_{i+1} |$. 
If all points are pairwise distinct, then $P$ is a \emph{path} in $K_S$. 
If the sequence may have repeated points, then $P$ is a \emph{walk} 
in $K_S$.  

\section{General Metric Spaces} 
\label{secGMS}
In this section, $(S,| \cdot |)$ denotes an arbitrary finite metric 
space. Let $n$ be the number of points in $S$ and let 
$t \geq 1$ be a real number. We assume that we are given a 
$t$-spanner $G'=(S,E')$ for $S$ and we denote the number of edges of 
$G'$ by $m$. For any integer $f$ with $1 \leq f \leq (n-1)/2$,  
we will show how to convert $G'$ to a graph $G=(S,E)$ that has at most 
$(4f-1) m$ edges and is an $f$-faulty-degree $t'$-spanner for $S$, 
where $t'$ is a function of $f$ and $t$.  

The graph $G=(S,E)$ is obtained as follows (refer to Figure~\ref{figE}): 

\begin{figure}[t]
\centering
\includegraphics[scale=1.2]{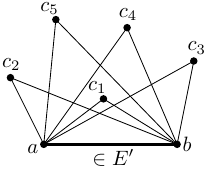}
\caption{Illustrating the edges that are added to $E$ because of the 
edge $\{a,b\}$ in $E'$. In this figure, $f=3$.} 
\label{figE}
\end{figure}

\begin{itemize}
\item The edge set $E$ contains all edges of $E'$. 
\item For each edge $\{a,b\}$ in $E'$, let $c_1,c_2,\ldots,c_{n-2}$ 
be the sequence of points of $S \setminus \{a,b\}$, in non-decreasing 
order of the values $|ac_i|+|c_i b|$; ties are broken arbitrarily. 
Let $C_{ab}$ denote the set consisting of the first $2f-1$ points in 
this sequence, i.e., $C_{ab} = \{ c_1,c_2,\ldots,c_{2f-1} \}$. The 
edge set $E$ contains the $4f-2$ edges $\{a,c_i\}$ and $\{c_i,b\}$, for 
$i=1,2,\ldots,2f-1$. 
\end{itemize}  
Observe that the number of edges in $E$ is at most 
$m + (4f-2)m = (4f-1)m$. 
 
In the rest of this section, we will prove that $G$ is an 
$f$-faulty-degree $t'$-spanner 
for $S$. In Section~\ref{secFM}, we consider the case when $f=1$, in 
which case $t' \leq 3t$. In Section~\ref{secgen}, we consider the 
general case when $f \geq 1$ and prove that $t' \leq (8f+2)t$.

\subsection{Faulty Matchings: $f=1$} 
\label{secFM} 

Consider the graph $G=(S,E)$ defined above for the case when $f=1$. 
Any subset $F$ of $E$ such that the maximum degree of the graph $(S,F)$ 
is at most $1$ is a matching in $G$; this matching may not be a perfect 
matching. In other words, $G \setminus F$ is 
obtained by removing this matching from $G$. We will prove the following 
result.  

\begin{theorem} 
\label{matching} 
Let $(S,| \cdot |)$ be a finite metric space, let $t \geq 1$ be a 
real number, and let $G'=(S,E')$ be a $t$-spanner for $S$ with $m$ 
edges. Then the graph $G=(S,E)$ is a $1$-faulty-degree 
$(3t)$-spanner for $S$ that has at most $3m$ edges. 
\end{theorem} 

We have already seen that the number of edges in $E$ is at most $3m$. 
It remains to prove that $G$ is a $1$-faulty-degree $(3t)$-spanner 
for $S$.

Let $F$ be an arbitrary matching in $G$, and let $p$ and $q$ be two 
points in $S$. We have to show that 
\[ \delta_{G \setminus F} (p,q) \leq 3t \cdot 
          \delta_{K_s \setminus F}(p,q) . 
\]

Let $P'$ be a shortest path between $p$ and $q$ in the graph $G'$. 
Since $G'$ is a $t$-spanner for $S$, the length $|P'|$ of $P'$ is at 
most $t |pq|$. 

\begin{figure}[t]
\centering
\includegraphics[scale=0.7]{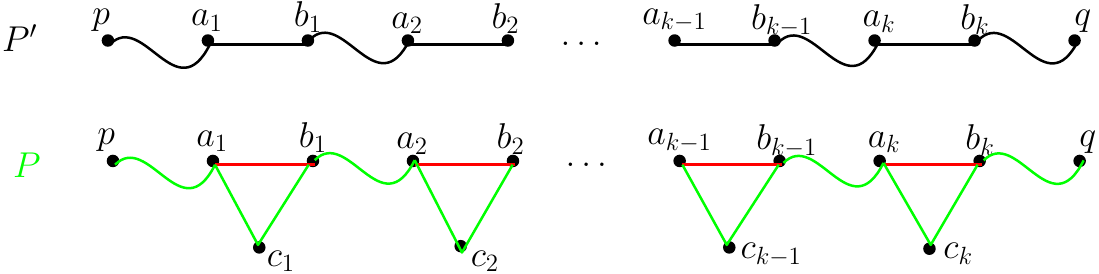}
\caption{$P'$ is a shortest path between $p$ and $q$ in the graph $G'$.
The red edges $\{ a_i,b_i \}$, $1 \leq i \leq k$, are in the matching $F$.
$P$ (in green) is a walk between $p$ and $q$ in the graph $G \setminus F$.
Note that any vertex $c_i$ may be on the path $P'$.}
\label{fig1}
\end{figure}

Let $\{a_1,b_1\}, \{a_2,b_2\} , \ldots , \{a_k,b_k\}$ be the edges on 
$P'$ that are in $F$. We may assume that, if we follow $P'$ from $p$ 
to $q$, the endpoints of these edges are visited in the order 
$a_1,b_1,a_2,b_2,\ldots,a_k,b_k$; refer to Figure~\ref{fig1}. Since 
$F$ is a matching, 
$b_1 \neq a_2 , b_2 \neq a_3 , \ldots , b_{k-1} \neq a_k$. 

For each $i$ with $1 \leq i \leq k$, let $c_i$ be the point in 
$S \setminus \{a_i,b_i\}$ that minimizes $|a_i c_i| + |c_i b_i|$ and 
that was chosen when defining the graph $G$. Since $F$ is a matching, 
all edges $\{a_i,c_i\}$ and $\{c_i,b_i\}$, for $1 \leq i \leq k$, are 
in $G \setminus F$. 

We define $P$ to be the walk (i.e., possibly non-simple path) that 
\begin{itemize}
\item starts at $p$ and follows $P'$ to $a_1$, 
\item for $i=1,2,\ldots,k-1$, takes the two edges $\{a_i,c_i\}$ and 
$\{c_i,b_i\}$, and then follows $P'$ from $b_i$ to $a_{i+1}$, 
\item takes the two edges $\{a_k,c_k\}$ and $\{c_k,b_k\}$, and then 
follows $P'$ from $b_k$ to $q$. 
\end{itemize}
Refer to Figure~\ref{fig1}. Observe that $P$ is a walk between $p$ and $q$ 
in the graph $G \setminus F$. 

If $k=0$, then $P = P'$ and we have 
\[ \delta_{G \setminus F} (p,q) = |P'| \leq t|pq| = 
     t \cdot \delta_{K_S \setminus F}(p,q) \leq 
     3t \cdot \delta_{K_S \setminus F}(p,q) .
\]  
Assume that $k=1$ and $P'$ consists of just the edge $\{p,q\}$. 
Then $p=a_1$, $q=b_1$, and the length of $P$ is equal to 
$|a_1 c_1| + |c_1 b_1|$. By the definition of $c_1$, we have 
$\delta_{K_S \setminus F}(p,q) = |a_1 c_1| + |c_1 b_1|$ and, thus, 
\[ \delta_{G \setminus F} (p,q) =  
   \delta_{K_S \setminus F}(p,q) \leq 
    3t \cdot \delta_{K_S \setminus F}(p,q) .  
\]
It remains to consider the case covered by the following lemma. 

\begin{lemma}
\label{claim1}
Assume that $k \geq 1$ and $P'$ has at least two edges. Then the 
graph $G \setminus F$ contains a walk $Q$ between $p$ and $q$ whose  
length is at most $3 |P'|$. 
\end{lemma} 

This lemma will imply that 
\[ \delta_{G \setminus F} (p,q) \leq |Q| \leq 3 |P'| \leq 3t |pq| \leq 
         3t \cdot \delta_{K_s \setminus F}(p,q) . 
\]

\subsubsection{Proof of Lemma~\ref{claim1}}
For any two vertices $x$ and $y$ of $P'$, we denote by $P'_{xy}$ the 
subpath of $P'$ between $x$ and~$y$. Using this notation, the length 
$|P|$ of $P$ satisfies 
\begin{equation} 
\label{eq1} 
  |P| = | P'_{p a_1} | + \sum_{i=1}^{k-1} | P'_{b_i a_{i+1}} | + 
   | P'_{b_k q} | + \sum_{i=1}^k \left( |a_i c_i| + |c_i b_i| \right) . 
\end{equation} 

To prove Lemma~\ref{claim1}, we consider four cases. In order to 
verify that these cases cover all possibilities, we describe them 
first: 

\begin{itemize}
\item Case 1: At least one of the two edges of $P'$ that are 
incident on $p$ and $q$ is not in $F$. 
\item In the remaining three cases, both edges of $P'$ that are 
incident on $p$ or $q$ are in $F$. 
\item Assume that $k=2$. Since $F$ is a matching in $G$, and $G'$ is 
a subgraph of $G$, $P'$ has at least three edges.  
\item Case 2: $k=2$ and $P'$ consists of three edges. 
\item Case 3: $k=2$ and $P'$ consists of at least four edges. 
\item Case 4: $k \geq 3$.
\end{itemize} 

\noindent 
{\bf Case 1:} The edge of $P'$ that is incident on $p$ is not in $F$
or the edge of $P'$ that is incident on $q$ is not in $F$. 

By symmetry, we may assume that the edge of $P'$ that is incident on $p$ 
is not in $F$.  
Since $p \not\in \{a_1,b_1\}$, our choice of $c_1$ implies that 
\[ |a_1 c_1| + |c_1 b_1| \leq |a_1 p| + |p b_1| . 
\]
For each $i$ with $2 \leq i \leq k$, since $b_{i-1} \not\in \{a_i,b_i\}$, 
our choice of $c_i$ implies that 
\[ |a_i c_i| + |c_i b_i| \leq |a_i b_{i-1}| + |b_{i-1} b_i| . 
\]
It follows that 

\begin{eqnarray*} 
  \sum_{i=1}^k \left( |a_i c_i| + |c_i b_i| \right) 
 & \leq & \left( |a_1 p| + |p b_1| \right) + 
          \sum_{i=2}^k \left( |a_i b_{i-1}| + |b_{i-1} b_i| \right) \\ 
 & \leq & |P'_{p a_1}| + |P'_{p b_1}| + 
          \sum_{i=2}^k \left( |P'_{b_{i-1} a_i}| + |P'_{b_{i-1} b_i}| 
                       \right) \\ 
 & \leq & 2 |P'| . 
\end{eqnarray*} 
Using this, it follows from (\ref{eq1}) that $|P| \leq 3|P'|$. Hence, 
Lemma~\ref{claim1} holds for $Q = P$. 

\vspace{0.5em}

\noindent 
{\bf Case 2:} $k=2$, the path $P'$ consists of three edges, 
the edge of $P'$ that is incident on $p$ is in $F$, and the edge of $P'$ 
that is incident on $q$ is in $F$. 

In this case, $p=a_1$, $q=b_2$, and the path $P$ consists of the five 
edges $\{ a_1,c_1 \}$, $\{ c_1,b_1 \}$, $\{ b_1,a_2 \}$, 
$\{ a_2,c_2 \}$, and $\{ c_2,b_2 \}$. Observe that $\{ b_1,a_2 \}$ is 
not in $F$. 

Let $c$ be the point in $S \setminus \{ b_1,a_2 \}$ that minimizes 
$|b_1 c| + |c a_2|$ and that was chosen when defining the graph $G$. 

\vspace{0.5em}

\noindent 
{\bf Case 2.1:} $c \not\in \{ a_1,b_2 \}$. 

Since $a_1 \not\in \{ b_1,a_2 \}$, our choice of $c$ implies that 
\[ | b_1 c | + | c a_2 | \leq | b_1 a_1 | + |a_1 a_2 | \leq 
      | b_1 a_1 | + | a_1 b_1 | + | b_1 a_2| . 
\] 
Similarly, since $b_2 \not\in \{ b_1,a_2 \}$, 
\[ | b_1 c | + | c a_2 | \leq | b_1 b_2 | + |b_2 a_2 | \leq 
    | b_1 a_2 | + | a_2 b_2 | + | b_2 a_2 | . 
\] 
By adding these two inequalities, we get 
\[ 2 \left( | b_1 c | + | c a_2 | \right) \leq 
   2 | a_1 b_1 | + 2 |b_1 a_2 | + 2 | a_2 b_2 |  = 2 |P'| ,
\]
which gives   
\begin{equation} 
\label{eq2} 
   | b_1 c | + | c a_2 | \leq |P'| .  
\end{equation} 

Since $c \not\in \{ a_1,b_1 \}$, our choice of $c_1$ implies that  
\begin{equation}
\label{eq3} 
   | a_1 c_1 | + | c_1 b_1 | \leq | a_1 c | + | c b_1 | . 
\end{equation} 
Since $c \not\in \{ a_2,b_2 \}$, our choice of $c_2$ implies that 
\begin{equation}
\label{eq4} 
   | a_2 c_2 | + | c_2 b_2 | \leq | a_2 c | + | c b_2 | . 
\end{equation} 
Using (\ref{eq3}) and (\ref{eq4}), we get 

\begin{eqnarray*} 
 |P| & = & | a_1 c_1 | + |c_1 b_1 | + | b_1 a_2 | + | a_2 c_2 | + 
           | c_2 b_2| \\ 
  & \leq & | a_1 c | + | c b_1 | + | b_1 a_2 | + | a_2 c | + | c b_2 | .
\end{eqnarray*} 
Using (\ref{eq2}), we get 

\begin{eqnarray*} 
 |P| & \leq & | a_1 c | + | b_1 a_2 | + | c b_2 | + |P'| \\ 
 & \leq & \left( | a_1 b_1 | + | b_1 c | \right) + | b_1 a_2 | + 
          \left( | c a_2 | + | a_2 b_2 | \right) + |P'| \\ 
 & = & | b_1 c | + | c a_2 | + 2 |P'| . 
\end{eqnarray*} 
One more application of (\ref{eq2}) gives 
\[ |P| \leq 3 |P'| . 
\] 
Hence, Lemma~\ref{claim1} holds for $Q = P$. 

\vspace{0.5em}

\noindent 
{\bf Case 2.2:} $c \in \{a_1,b_2\}$. 

By symmetry, we may assume that $c=a_1$. 
By the definition of the graph $G$, $\{ c,a_2 \} = \{ p,a_2 \}$ is 
an edge in $G$. Since $F$ is a matching, this edge is in the graph 
$G \setminus F$. Define the path $Q = (p,a_2,c_2,b_2=q)$. This 
path is in $G \setminus F$. 

Since $c \not\in \{ a_2,b_2 \}$, our choice of $c_2$ implies that 
\[ | a_2 c_2 | + | c_2 b_2 | \leq | a_2 c | + | c b_2 | , 
\]
i.e., 
\[ | a_2 c_2 | + | c_2 q | \leq | a_1 a_2 | + | a_1 b_2 | . 
\]
The length of the path $Q$ satisfies 
\[ |Q| = | p a_2 | + | a_2 c_2 | + | c_2 q | \leq 
      | p a_2 | + | a_1 a_2 | + | a_1 b_2 | . 
\]
By the triangle inequality, we have $|Q| \leq 3 |P'|$. Hence, 
Lemma~\ref{claim1} holds for $Q$. 

\vspace{0.5em}

\noindent 
{\bf Case 3:} $k=2$, the path $P'$ consists of at least four edges, 
the edge of $P'$ that is incident on $p$ is in $F$, and the edge of $P'$ 
that is incident on $q$ is in $F$. 

In this case, $p=a_1$ and $q=b_2$. Let $z$ be the third point on $P'$, 
when traversing this path from $p$ to $q$. Then, the edge 
$\{ b_1,z \}$ is on $P'$ and $z \neq a_1$. 

Observe that $P'_{pz}$ is a shortest path between $p$ and $z$ in $G'$ and 
this path satisfies the condition of Case~2. Therefore, the graph 
$G \setminus F$ contains a walk $Q_1$ between $p$ and $z$ whose length 
is at most $3 | P'_{pz} |$. 

Similarly, $P'_{zq}$ is a shortest path between $z$ and $q$ in $G'$ and 
this path satisfies the condition of Case~1. Thus, the graph 
$G \setminus F$ contains a walk $Q_2$ between $z$ and $q$ whose length 
is at most $3 |P'_{zq} |$. 

Let $Q$ be the concatenation of $Q_1$ and $Q_2$. Then $Q$ is a walk 
between $p$ and $q$ in $G \setminus F$. The length of this walk 
satisfies  
\[ |Q| = | Q_1 | + | Q_2 | \leq 3 | P'_{pz} | + 3 | P'_{zq} | = 
               3 |P'| . 
\] 
Thus, Lemma~\ref{claim1} holds for this walk $Q$. 

\vspace{0.5em}

\noindent 
{\bf Case 4:} $k \geq 3$, the edge of $P'$ that is incident on $p$ is 
in $F$, and the edge of $P'$ that is incident on $q$ is in $F$. 

In this case, $p = a_1$ and $q = b_k$. Observe that $P'_{p b_2}$ is a 
shortest path between $p$ and $b_2$ in $G'$ and this path satisfies one 
of the conditions of Cases~3 and~4. Therefore, the graph $G \setminus F$
contains a walk $Q_1$ between $p$ and $b_2$ whose length is at most 
$3 | P'_{p b_2} |$. Similarly, $P'_{b_2 q}$ is a shortest path between 
$b_2$ and $q$ in $G'$ and this path satisfies the condition of Case~1. 
Thus, the graph $G \setminus F$ contains a walk between $b_2$ and $q$ 
whose length is at most $3 | P'_{b_2 q} |$. Let $Q$ be the concatenation 
of $Q_1$ and $Q_2$. Then $Q$ is a walk between $p$ and $q$ in 
$G \setminus F$. The length of this walk satisfies 
\[ |Q| = | Q_1 | + | Q_2 | \leq 3 | P'_{p b_2} | + 3 | P'_{b_2 q} | 
          = 3 |P'| . 
\] 
Thus, Lemma~\ref{claim1} holds for this walk $Q$. 

\vspace{0.5em}

This completes the proof of Lemma~\ref{claim1} and, therefore, 
of Theorem~\ref{matching}.  

\subsubsection{Lower Bound} 
We show that the upper bound of $3t$ on the stretch factor of the 
graph $G$ in Theorem~\ref{matching} is tight. 
Let $\eps$ be a real number with $0 < \eps \ll 1$, and define 
$\eps' = \eps / (4 - 2 \eps)$. Consider the one-dimensional point set 
$S = \{p,a,b,q\}$, as indicated below. The graph $G'$ connects the 
points in the left-to-right order and is a $1$-spanner for $S$. The 
graph $G$ contains $G'$ and the two edges $\{p,b\}$ and $\{a,q\}$.  

\begin{center}
\includegraphics[scale=0.7]{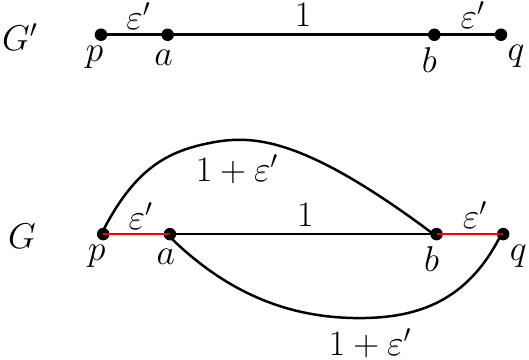}
\end{center}

Consider the matching $F = \{ \{p,a\} , \{b,q\} \}$, as indicated in red. 
We have $\delta_{G \setminus F}(p,q) = 3 + 2 \eps'$ and 
$\delta_{K_S \setminus F}(p,q) = 1 + 2 \eps'$. Thus, 
\[ \frac{\delta_{G \setminus F}(p,q)}{\delta_{K_S \setminus F}(p,q)} 
     = \frac{3 + 2 \eps'}{1 + 2 \eps'} = 3 - \eps ,  
\] 
i.e., $G'$ is a $1$-faulty-degree $(3-\eps)$-spanner for $S$.

\subsection{The General Case: $f \geq 1$} 
\label{secgen} 

Consider the graph $G=(S,E)$ defined in the beginning of 
Section~\ref{secGMS} for the general case when $f \geq 1$. 
We will prove the following result.  

\begin{theorem} 
\label{thmgen} 
Let $(S,| \cdot |)$ be a finite metric space, let $n$ be the number of 
points in $S$, let $t \geq 1$ be a real number, and let $G'=(S,E')$ be a 
$t$-spanner for $S$ with $m$ edges. Then, for any integer $f$ with 
$1 \leq f \leq (n-1)/2$,  the graph $G=(S,E)$ is an $f$-faulty-degree 
$((8f+2)t)$-spanner for $S$ that has at most $(4f-1) m$ edges. 
\end{theorem} 

We have already seen that the number of edges in $E$ is at most 
$(4f-1)m$. It remains to prove that $G$ is an 
$f$-faulty-degree $((8f+2)t)$-spanner for $S$.

Let $F$ be an arbitrary subset of the edge set $E$ of $G$, such that
the maximum degree of the graph $(S,F)$ is at most $f$. 
In the rest of this section, we will prove that for any two distinct 
points $p$ and $q$ in $S$, 
\begin{equation} 
\label{toprove} 
   \delta_{G \setminus F}(p,q) \leq 
             (8f+2)t \cdot \delta_{K_S \setminus F}(p,q) . 
\end{equation} 
Since the proof is long and quite intricate, we have broken it up into  
several preliminary lemmas. The proof of (\ref{toprove}) will be given
in Lemma~\ref{C5}. 

\begin{lemma}
\label{C0} 
The graph $K_S \setminus F$ is connected. 
\end{lemma} 
\begin{proof}
We will prove the following claim, which will imply the statement in 
the lemma: For every partition of 
$S$ into two non-empty sets $A$ and $B$, the graph $K_S \setminus F$ 
contains an edge between $A$ and $B$. 

To prove this claim, we may assume without loss of generality that 
the size of $B$ is at least the size of $A$. Thus, the size of $B$ 
is at least $n/2$, which is larger than $f$. Let $p$ be a point in $A$. 
In $K_S$, there are more than $f$ edges of the form $\{p,q\}$, 
with $q \in B$. At least one these edges is in $K_S \setminus F$. 
\end{proof}

We will repeatedly use the following lemma, which applies to any 
edge in $G'$ that is in the matching $F$. 

\begin{lemma}
\label{C1} 
Let $\{a,b\}$ be an edge in $G'$. If this edge is in $F$, then there 
exists a point $c$ in $C_{ab}$ such that both $\{a,c\}$ and $\{c,b\}$ are 
edges in the graph $G \setminus F$. 
\end{lemma} 
\begin{proof}
Assume that for each of the $2f-1$ points $c$ in $C_{ab}$, at least one 
of the two edges $\{a,c\}$ and $\{c,b\}$ is in $F$. Then, in $(S,F)$, the 
sum of the degrees of $a$ and $b$ is at least $2 + (2f-1) = 2f+1$. 
Therefore, $a$ or $b$ has degree more than $f$ in $(S,F)$, which is a 
contradiction.  
\end{proof} 

The next lemma states that for any edge $\{a,b\}$ in $G'$, the graph 
$G \setminus F$ contains a shortest path, all of whose edges are 
``close'' to both $a$ and $b$. 

\begin{figure}[t]
\centering
\includegraphics[scale=1.2]{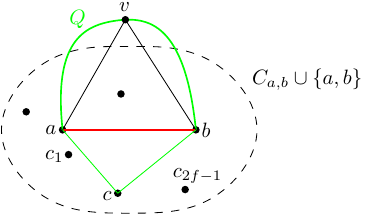}
\caption{Illustrating the proof of Lemma~\ref{C2}.} 
\label{figCab}
\end{figure}

\begin{lemma}
\label{C2} 
Let $\{a,b\}$ be an edge in $G'$. Then there exists a shortest path 
between $a$ and $b$ in the graph $G \setminus F$, all of whose vertices
belong to $C_{ab} \cup \{a,b\}$. 
\end{lemma} 
\begin{proof}
First assume that $\{a,b\}$ is not an edge in $F$. Then this edge is in 
the graph $G \setminus F$. Using the triangle inequality, this edge is a 
shortest path between $a$ and $b$. Thus, the lemma holds. 

Now assume that $\{a,b\}$ is in $F$. Let $c$ be a point in $C_{ab}$ as 
in Lemma~\ref{C1}. The path $P = (a,c,b)$ is in $G \setminus F$. 
Let $Q$ be a shortest path between $a$ and $b$ in $G \setminus F$, and 
assume that $Q$ contains a vertex $v$ that is not in 
$C_{ab} \cup \{a,b\}$; refer to Figure~\ref{figCab}. Then 
\[ |P| = |ac|+|cb| \leq |av|+|vb| \leq |Q| 
\]
and, thus, $|P|=|Q|$. The path $P$ satisfies the lemma. 
\end{proof} 

The next lemma states that, in the graph $G \setminus F$, any two points 
$p$ and $q$ are connected by a ``short'' path, if their shortest path 
in $G'$ contains ``many'' edges. 
 
\begin{lemma}
\label{C4} 
Let $p$ and $q$ be two distinct points in $S$, and let $P'$ be a shortest 
path between $p$ and $q$ in the graph $G'$. If the number of vertices 
of $P'$ is at least $2f+2$, then 
\[ \delta_{G \setminus F}(p,q) \leq (8f+2) \cdot |P'| . 
\] 
\end{lemma} 
\begin{proof}
Let $e=\{a,b\}$ be any edge on the path $P'$. By Lemma~\ref{C2}, there 
exists a shortest path $P_e$ between $a$ and $b$ in the graph 
$G \setminus F$, all of whose vertices belong to $C_{ab} \cup \{a,b\}$;
refer to Figure~\ref{figP}. 

\begin{figure}[t]
\centering
\includegraphics[scale=1.2]{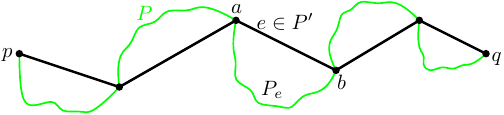}
\caption{The black path $P'$ is a shortest path between $p$ and $q$ 
in $G'$. For each edge $e=\{a,b\}$ on $P'$, the green path $P_e$ is a 
shortest path between $a$ and $b$ in $G \setminus F$. The concatenation of 
all paths $P_e$ is a walk $P$ between $p$ and $q$ in $G \setminus F$.} 
\label{figP}
\end{figure}

Let $P$ be the concatenation of all paths $P_e$, where $e$ ranges over 
all edges of $P'$. Then $P$ is a walk between $p$ and $q$ in 
$G \setminus F$. We will prove that the length of $P$ is at most 
$(8f+2) \cdot |P'|$. This will prove the lemma. 

\begin{figure}[t]
\centering
\includegraphics[scale=1.2]{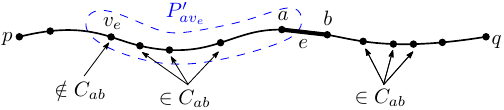}
\caption{$v_e$ is a vertex of $P'$ that is closest (along $P'$ and 
in terms of the number of edges) to $a$ or $b$, and that is not in 
$C_{ab} \cup \{a,b\}$.} 
\label{figPave}
\end{figure}

Again, let $e=\{a,b\}$ be any edge on the path $P'$. Recall that the 
set $C_{ab} \cup \{a,b\}$ has size $2f+1$ and $P'$ contains at least 
$2f+2$ vertices. Therefore, there exists a vertex of $P'$ that is not 
in $C_{ab} \cup \{a,b\}$. Let $v_e$ be such a vertex that is closest 
(along $P'$ and in terms of the number of edges), to $a$ or $b$; refer 
to Figure~\ref{figPave}. Then, again using the fact that the size of 
$C_{ab} \cup \{a,b\}$ is $2f+1$,  
\begin{itemize} 
\item the subpath $P'_{a v_e}$ of $P'$ between $a$ and $v_e$ has at most 
$2f+1$ edges, and  
\item the subpath $P'_{v_e b}$ of $P'$ between $b$ and $v_e$ has at most 
$2f+1$ edges. 
\end{itemize} 
First assume that $e = \{a,b\}$ is an edge in $F$. Let $c$ be a point 
as in Lemma~\ref{C1}. Since both edges $\{a,c\}$ and $\{c,b\}$ are in 
$G \setminus F$, we have 
\[ \delta_{G \setminus F}(a,b) \leq |ac|+|cb| .
\] 
Since $v_e \not\in C_{ab}$, we have 
\[ |ac|+|cb| \leq |a v_e| + |v_e b| . 
\] 
By the triangle inequality, we have 
\[ |a v_e| + |v_e b| \leq |P'_{a v_e} | + |P'_{v_e b}| .
\] 
By combining the above three inequalities, we get 
\begin{equation} 
\label{eq66} 
   \delta_{G \setminus F}(a,b) \leq |P'_{a v_e} | + |P'_{v_e b}| .
\end{equation} 
Observe that (\ref{eq66}) also holds if $e=\{a,b\}$ is not an edge in 
$F$, because in that case, $\delta_{G \setminus F}(a,b) = |ab|$, which 
is at most equal to the right-hand side in (\ref{eq66}).  

Since $P$ is the concatenation of all paths $P_e$, with $e$ an edge 
of $P'$, we have 
\begin{eqnarray} 
 |P| & = & \sum_{e=\{a,b\} \mbox{\scriptsize{ on }} P'} |P_e| 
               \nonumber \\ 
 & = & \sum_{e=\{a,b\} \mbox{\scriptsize{ on }} P'} 
                 \delta_{G \setminus F}(a,b) \nonumber \\ 
 & \leq & \sum_{e=\{a,b\} \mbox{\scriptsize{ on }} P'} 
          \left( |P'_{a v_e} | + |P'_{v_e b}| \right) \nonumber \\  
 & = & \sum_{e=\{a,b\} \mbox{\scriptsize{ on }} P'} |P'_{a v_e} | + 
       \sum_{e=\{a,b\} \mbox{\scriptsize{ on }} P'} |P'_{v_e b}| .
               \label{RHS}  
\end{eqnarray} 

Consider the first summation in (\ref{RHS}). For each edge $\{x,y\}$ 
on the path $P'$, let $N_{xy}$ be the number of edges $e=\{a,b\}$ 
on $P'$ such that $\{x,y\}$ is an edge on the subpath $P'_{a v_e}$. Then 
\[ \sum_{e=\{a,b\} \mbox{\scriptsize{ on }} P'} |P'_{a v_e} | = 
   \sum_{\{x,y\} \mbox{\scriptsize{ on }} P'} N_{xy} \cdot |xy| . 
\] 
To obtain an upper bound on $N_{xy}$, we assume, without loss of 
generality, that $x$ is on the subpath of $P'$ between $p$ and $y$. 
Define the following subpath $Q_{xy}$ of $P'$:
\begin{itemize}
\item $\{x,y\}$ is an edge of $Q_{xy}$. 
\item Starting at $x$, traverse $P'$ towards $p$ and stop as soon as 
$2f$ edges have been traversed or the point $p$ has been reached.  
All edges traversed are edges of $Q_{xy}$. 
\item Starting at $y$, traverse $P'$ towards $q$ and stop as soon as 
$2f$ edges have been traversed or the point $q$ has been reached.  
All edges traversed are edges of $Q_{xy}$. 
\end{itemize} 
Recall that each subpath $P'_{a v_e}$ has at most $2f+1$ edges. 
It follows that every edge $e=\{a,b\}$ that is counted in $N_{xy}$ 
is an edge on $Q_{xy}$. Since $Q_{xy}$ has at most $4f+1$ edges, it 
follows that $N_{xy} \leq 4f+1$. Thus, 
\[ \sum_{e=\{a,b\} \mbox{\scriptsize{ on }} P'} |P'_{a v_e} | \leq  
   (4f+1) \sum_{\{x,y\} \mbox{\scriptsize{ on }} P'} |xy| = 
   (4f+1) \cdot |P'| . 
\] 
By a symmetric argument, the second summation in (\ref{RHS}) is at most 
$(4f+1) \cdot |P'|$. We conclude that $|P| \leq (8f+2) |P'|$. 
\end{proof} 

We are now ready to complete the proof of Theorem~\ref{thmgen}. 
The proof of the next lemma uses the previous lemmas to obtain an 
upper bound on the length of a shortest path in $G \setminus F$ 
between any pairs of points.   

\begin{lemma}
\label{C5} 
Let $p$ and $q$ be two distinct points in $S$. Then 
\[ \delta_{G \setminus F}(p,q) \leq 
             (8f+2)t \cdot \delta_{K_S \setminus F}(p,q) . 
\] 
\end{lemma} 
\begin{proof}
Let $Q$ be a shortest path between $p$ and $q$ in $K_S \setminus F$. 
By Lemma~\ref{C0}, $Q$ exists. 
For each edge $e=\{a,b\}$ on $Q$, let $P'_e$ be a shortest path between 
$a$ and $b$ in $G'$. Since $G'$ is a $t$-spanner for $S$, we have 
$|P'_e| \leq t|ab|$. We will show that 

\begin{equation}
\label{eqstar} 
  \delta_{G \setminus F}(a,b) \leq (8f+2) \cdot |P'_e| . 
\end{equation}
This will imply that 

\begin{eqnarray*} 
  \delta_{G \setminus F}(p,q) & \leq & 
  \sum_{e=\{a,b\} \mbox{\scriptsize{ on }} Q} 
           \delta_{G \setminus F}(a,b) \\ 
 & \leq & \sum_{e=\{a,b\} \mbox{\scriptsize{ on }} Q} 
                    (8f+2) \cdot |P'_e| \\ 
 & \leq & \sum_{e=\{a,b\} \mbox{\scriptsize{ on }} Q} 
                    (8f+2)t \cdot |ab| \\ 
 & = & (8f+2)t \cdot |Q| \\ 
 & = & (8f+2)t \cdot \delta_{K_S \setminus F}(p,q) .  
\end{eqnarray*} 

If $P'_e$ has at least $2f+2$ vertices, then (\ref{eqstar}) follows 
from Lemma~\ref{C4}. From now on, we assume that $P'_e$ has at most 
$2f+1$ vertices and, thus, at most $2f$ edges. 

Since $Q$ is a path in $K_S \setminus F$ and $e=\{a,b\}$ is an edge on 
$Q$, this edge is not in $F$. If $e$ is an edge in $G \setminus F$, 
then 
\[ \delta_{G \setminus F}(a,b) = |ab| \leq |P'_e| 
\]
and, thus (\ref{eqstar}) holds.  

It remains to consider the case when $e=\{a,b\}$ is not an edge in 
$G \setminus F$. Consider the edge $\{a,d\}$ on $P'_e$ that is incident 
on $a$. Observe that $d \neq b$, because $\{a,b\}$ is not an edge in 
$G'$. 

We first show that for every point $z$ in $C_{ad}$, 
\begin{equation} 
\label{eqhello} 
  |az| + |zd| \leq |ab| + |bd| . 
\end{equation} 
The proof is by contradiction: If $|ab| + |bd| < |az| + |zd|$, then 
$b$ is an element of $C_{ad}$ and, therefore, by the definition of 
the graph $G$, $\{a,b\}$ is an edge in $G$, which is a contradiction. 

Next, we show that 
\begin{equation} 
\label{eqhello2} 
  \delta_{G \setminus F}(a,d) \leq 2 \cdot |P'_e| . 
\end{equation} 
If $\{a,d\}$ is not in $F$, then 
$\delta_{G \setminus F}(a,d) = |ad| \leq |P'_e|$ and, thus, 
(\ref{eqhello2}) holds. Assume that $\{a,d\}$ is an edge in $F$. By 
Lemma~\ref{C1}, there exists a point $c$ in $C_{ad}$, such that both 
$\{a,c\}$ and $\{c,d\}$ are edges in $G \setminus F$. Then, 
$\delta_{G \setminus F}(a,d) \leq |ac|+|cd|$, which, by (\ref{eqhello}), 
is at most $|ab|+|bd|$, which is at most $2 \cdot | P'_e |$, i.e., 
(\ref{eqhello2}) holds. 

Let $D = C_{ad} \cup \{a,d\}$. Observe that $D$ has size $2f+1$. 
Let $\{x,y\}$ be an arbitrary edge on $P'_e$ with 
$\{x,y\} \neq \{a,d\}$. Let $D' = D \setminus \{x,y\}$. Then the size 
of $D'$ is at least $2f-1$. Let $d'$ be a point in $D'$ for which 
$|xd'|+|d'y|$ is maximum. Refer to Figure~\ref{figCxy-2}. 

\begin{figure}[t]
\centering
\includegraphics[scale=1.2]{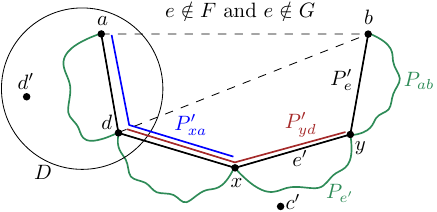}
\caption{$D = C_{ad} \cup \{a,d\}$, $\{x,y\}$ is an edge on $P'_e$, 
$\{x,y\} \neq \{a,d\}$.} 
\label{figCxy-2}
\end{figure}

First, we assume that $\{x,y\}$ is an edge in $F$. By Lemma~\ref{C1}, 
there is a point $c'$ in $C_{xy}$ such that both $\{x,c'\}$ and 
$\{c',y\}$ are edges in $G \setminus F$. It is clear that 
\[ \delta_{G \setminus F}(x,y) \leq |xc'| + |c'y| . 
\] 
We next argue that 
\[ |xc'|+|c'y| \leq |xd'|+|d'y| . 
\] 
Assume, by contradiction, that $|xd'|+|d'y| < |xc'|+|c'y|$. Then for 
all $z$ in $D' = D \setminus \{x,y\}$, $|xz|+|zy| < |xc'|+|c'y|$. 
There are at least $2f-1$ such points $z$, and each of them is in 
$S \setminus \{x,y\}$. Thus, $c' \not\in C_{xy}$, which is a 
contradiction.  

The previous two inequalities, together with the triangle inequality, 
imply that 
\begin{eqnarray*} 
  \delta_{G \setminus F}(x,y) \leq |xd'| + |d'y| 
  & \leq & \left( |P'_{xa}| + |ad'| \right) + 
          \left( |d'd| + |P'_{dy}|  \right) \\ 
  & \leq & |ad'|+|d'd| + 2 \cdot |P'_e| . 
\end{eqnarray*} 

We claim that 
\begin{equation} 
\label{eqhello3} 
  |ad'|+|d'd| \leq |ab|+|bd| . 
\end{equation} 
If $d' \in \{a,d\}$, then (\ref{eqhello3}) follows from the triangle 
inequality. If $d' \not\in \{a,d\}$, then $d' \in C_{ad}$ and 
(\ref{eqhello3}) follows from (\ref{eqhello}). 

Since $|ab|+|bd| \leq 2 \cdot |P'_e|$, we conclude that 
\begin{equation} 
\label{eqhello4} 
  \delta_{G \setminus F}(x,y) \leq 4 \cdot |P'_e| . 
\end{equation} 
In the derivation of (\ref{eqhello4}), we assumed that $\{x,y\}$ is an 
edge in $F$. If $\{x,y\}$ is not an edge in $F$, then (\ref{eqhello4}) 
also holds, because, in that case, 
$\delta_{G \setminus F}(x,y) = |xy| \leq |P'_e|$. 
   
%\begin{figure}[t]
%\centering
%\includegraphics[scale=1.2]{Cxy}
%\caption{The black path $P'_e$ is a shortest path between $a$ and $b$ 
%in $G'$. For each edge $e'=\{x,y\}$ on $P'_e$, the green path $P_{e'}$ 
%is a shortest path between $x$ and $y$ in $G \setminus F$. The 
%concatenation of all paths $P_{e'}$ is a walk $P_{ab}$ between $a$ and 
%$b$ in $G \setminus F$.} 
%\label{figCxy}
%\end{figure}

For each edge $e'=\{x,y\}$ on $P'_e$, let $P_{e'}$ be a shortest path 
between $x$ and $y$ in $G \setminus F$. 
Let $P_{ab}$ be the concatenation of all paths $P_{e'}$, where $e'$ 
ranges over all edges of $P'_e$. Observe that $P_{ab}$ is a walk between 
$a$ and $b$ in $G \setminus F$. We have 
\begin{eqnarray*} 
 \delta_{G \setminus F}(a,b) & \leq & |P_{ab}| \\ 
  & = & \sum_{e'=\{x,y\} \mbox{\scriptsize{ on }} P'_e} |P_{e'}| \\ 
  & = & \sum_{e'=\{x,y\} \mbox{\scriptsize{ on }} P'_e} 
            \delta_{G \setminus F}(x,y) \\ 
  & \leq & \sum_{e'=\{x,y\} \mbox{\scriptsize{ on }} P'_e} 
            4 \cdot |P'_e| . 
\end{eqnarray*} 
Since $P'_e$ has at most $2f$ edges, the latter summation has at most 
$2f$ terms. We conclude that 
\[ \delta_{G \setminus F}(a,b) \leq 8f \cdot |P'_e| 
\]
and, thus, (\ref{eqstar}) holds.  
\end{proof} 

\subsubsection{A Lower Bound for our Construction} 
We have shown that any $t$-spanner $G'$ can be transformed to an 
$f$-faulty-degree $t'$-spanner with $t' = O(ft)$. Below, we will give 
an example for which $t' = \Omega(ft)$. 

Let $f \geq 3$ be an integer and let $\eps$ be a real number with 
$0 < \eps \ll 1/(f-2)$. Let $B_0,B_1,\ldots B_f$ be pairwise disjoint 
sets, where $B_0 = \{ b_0,b_1,\ldots,b_f\}$, and each $B_i$, for 
$1 \leq i \leq f$, consists of $2f$ points. Let $S$ be the union of 
$B_0,B_1,\ldots B_f$. 

We define the following graph $H$ with vertex set $S$; refer to 
Figure~\ref{figLBf}: 

\begin{figure}[t]
\centering
\includegraphics[scale=0.8]{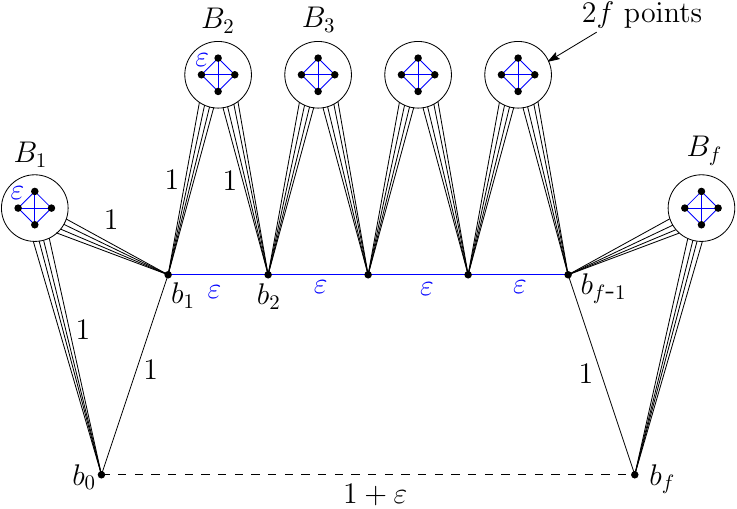}
\caption{Illustrating the graph $H$. All black edges have weight $1$, 
all blue edges have weight $\eps$, and the dotted edge $\{b_0,b_f\}$ has 
weight $1+\eps$. The set $B_0$ is equal to $\{b_0,b_1,\ldots,b_f\}$. 
The graph $G'$ consists of all solid edges.} 
\label{figLBf}
\end{figure}

\begin{itemize}
\item $H$ contains the edge $\{b_0,b_f\}$ having weight $1+\eps$. 
\item $H$ contains the edges $\{b_0,b_1\}$ and $\{b_{f-1},b_f\}$, each 
having weight $1$. 
\item $H$ contains the edges $\{b_j,b_{j+1}\}$, for $j=1,2,\ldots,f-2$, 
each having weight $\eps$. 
\item For each $i$ with $1 \leq i \leq f$, $H$ contains the complete 
graph on $B_i$, with each edge having weight $\eps$. 
\item For each $i$ with $1 \leq i \leq f$, $H$ contains the complete 
bipartite graph on $B_i \times \{b_{i-1},b_i\}$, with each edge 
having weight $1$.  
\end{itemize}

For any two points $p$ and $q$ of $S$, we define $|pq|$ to be the 
shortest-path distance between $p$ and $q$ in the graph $H$. It is clear 
that $(S,| \cdot |)$ is a metric space. 

\begin{observation}
\label{obs1} 
For each edge $\{p,q\}$ in $H$, $|pq|$ is the weight of this edge, 
i.e., $\{p,q\}$ is the shortest path between $p$ and $q$ in $H$. 
\end{observation}
\begin{proof}
The proof can be obtained by verifying that for each edge $\{p,q\}$,
any path between $p$ and $q$ that has at least two edges has length 
more than the weight of $\{p,q\}$. 
\end{proof} 

Let $G'$ be the graph obtained from $H$ by removing the edge 
$\{b_0,b_f\}$.

\begin{observation}
\label{obs2} 
$G'$ is a $3$-spanner for $S$. 
\end{observation}
\begin{proof}
If $\{x,y\}$ is an edge of $H$ and $\{x,y\} \neq \{b_0,b_f\}$, then,
by Observation~\ref{obs1}, $\delta_{G'}(x,y) = |xy|$. For the edge 
$\{b_0,b_f\}$, we have, using the fact that $\eps < 1/(f-2)$, 
\[ \delta_{G'}(b_0,b_f) = 1 + (f-2) \eps + 1 \leq 3 \leq  
    3 \cdot |b_0 b_f| .
\]
It follows that, for any two points $p$ and 
$q$ in $S$, 
\[ \delta_{G'}(p,q) \leq 3 \cdot \delta_{H}(p,q) = 3 \cdot |pq| . 
\] 
\end{proof} 

Having defined the $3$-spanner $G'$, consider the graph $G$ obtained 
from $G'$ by applying the transformation given in the beginning of
Section~\ref{secgen}. 

\begin{observation}
\label{obs3} 
$\{b_0,b_f\}$ is not an edge in the graph $G$.
\end{observation}
\begin{proof}
Let $\{p,q\}$ be any edge in $G'$. Recall that $G$ contains the complete 
graph with vertex set $C_{pq} \cup \{p,q\}$. The claim holds if we can 
show that at least one of $b_0$ and $b_f$ is not in 
$C_{pq} \cup \{p,q\}$. 

Observe that $|p b_0| +|b_0 q| \geq 2+\eps$ or 
$|p b_f|+|b_f q| \geq 2+\eps$, while there are at least $2f$ points $x$ 
in $S \setminus \{p,q\}$ such that $|px|+|xq| \leq 2$. This implies 
that at least one of $b_0$ and $b_f$ is not in $C_{pq} \cup \{p,q\}$.
\end{proof} 

\begin{observation}
\label{obs4} 
For any $i$ and $j$ with $1 \leq i < j \leq f$, the graph $G$ does 
not contain any edge between the sets $B_i$ and $B_j$. 
\end{observation}
\begin{proof}
For any point $a$ in $B_i$ and any point $b$ in $B_j$, 
we have $|ab| \geq 2$. 
For any edge $\{p,q\}$ in $G'$, we have $|pa|+|aq| \geq 2+\eps$ or 
$|pb|+|bq| \geq 2+\eps$, while there are at least $2f$ points $x$ in 
$S \setminus \{p,q\}$ such that $|px|+|xq| \leq 2$. This implies that 
at least one of $a$ and $b$ is not in $C_{pq} \cup \{p,q\}$.
\end{proof} 

By Theorem~\ref{thmgen}, the graph $G$ is an $f$-faulty-degree 
$t'$-spanner for $S$, where $t' \leq 3(8f+2) = 24f+6$. Below, we show 
that $t' \geq f$.  

Let $F$ be the set of all edges $\{p,q\}$ of $G$ for which both $p$ 
and $q$ are in $B_0$. Since $B_0$ has size $f+1$, the maximum degree of 
the graph $(S,F)$ is at most $f$. 

By Observation~\ref{obs3}, $\{b_0,b_f\}$ is not an edge in 
$G \setminus F$. This, together with Observation~\ref{obs4}, implies
that the shortest path between $b_0$ and $b_f$ in 
$G \setminus F$ consists of the following: 
\begin{itemize}
\item For $i=0,1,\ldots,f-1$, it takes the edge from $b_i$ to an 
arbitrary point, say $x_{i+1}$, of $B_{i+1}$, and then the edge 
from $x_{i+1}$ to $b_{i+1}$.  
\end{itemize}
Thus, 
\[ \delta_{G \setminus F}(b_0,b_f) = 2f . 
\] 
Since by Observation~\ref{obs3}, $\{b_0,b_f\}$ is not in $F$, we have 
\[ \delta_{K_S \setminus F}(b_0,b_f) = | b_0 b_f | = 1 + \eps 
            \leq 2 . 
\] 
It follows that 
\[ \frac{\delta_{G \setminus F}(b_0,b_f)}
        {\delta_{K_S \setminus F}(b_0,b_f)} \geq f .  
\]

\subsubsection{Comparison with the Bodwin--Haeupler--Parter 
Construction} 

Recall that Bodwin, Haeupler, and Parter~\cite{bhp-23} proved the 
following: Let $H$ be any edge-weighted graph with $n$ vertices. For any 
positive integers $f$ and $k$, there exists an $f$-faulty-degree 
$(2k-1)$-spanner for $H$ that has 
\[ O\left( k^k \cdot f^{1-1/k} \cdot n^{1+1/k} \right)
\]
edges. In particular, this results holds when $H$ is the complete 
graph induced by a metric space with $n$ points. 

Let $S$ be an arbitrary metric space with $n$ points. For any 
integer $k' \geq 1$, the greedy algorithm of 
Alth{\"o}fer \emph{et al.}~\cite{addjs-sswg-93} constructs a 
$(2k'-1)$-spanner for $S$ with $O(n^{1+1/k'})$ eges. 
By taking $k' = \Theta(k/f)$ and applying Theorem~\ref{thmgen} to this 
spanner, we obtain an $f$-faulty-degree $O(k)$-spanner for $S$, 
whose number of edges is 
\[ O\left( f \cdot n^{1+1/k'} \right) = 
   n^{\Theta(f/k)} \cdot f \cdot n^{1+1/k} . 
\]

If $k = \omega(\sqrt{f \cdot \log n})$, then 
$n^{\Theta(f/k)} = o( k^k )$ and, thus, the number of edges in our 
construction is smaller than the number of edges in the 
Bodwin--Haeupler--Parter construction.

\section{Metric Spaces with a Small WSPD} 
\label{secWSPD}

The construction in the previous section is valid for any metric space
that admits a $t$-spanner, for some value of $t$. However, the stretch 
factor of the $f$-faulty-degree spanner is $\Theta(ft)$. In this section, 
we will show that, if the metric space admits a well-separated pair 
decomposition, we can obtain a stretch factor of $1+\eps$, for any 
given $\eps>0$. 

\subsection{Preliminaries} 

Let $(S,|\cdot|)$ be a finite metric space. For any two non-empty
subsets $A$ and $B$ of $S$, we define their distance $|AB|$ as
\[ |AB| = \min \{ |pq| : p \in A , q \in B \} 
\]
and the diameter $\diam(A)$ of $A$ as
\[ \diam(A) = \max \{ |pq| : p \in A , q \in A \} . 
\]
For a real number $c>0$, called the \emph{separation ratio}, we say
that $A$ and $B$ are \emph{well-separated}, if
\[ |AB| \geq c \cdot \max ( \diam(A) , \diam(B) ) .
\]

\begin{lemma}  \label{4insamepair}
Let $c>0$ be a real number, let $A$ and $B$ be two non-empty subsets 
of $S$ that are well-separated with respect to $c$, let $p$ and $p'$ be 
any two points in $A$, and let $q$ and $q'$ be any two points in $B$. 
Then 
\begin{enumerate}
\item $|pp'| \leq (1/c) |pq|$, 
\item $|qq'| \leq (1/c) |pq|$, and 
\item $|p'q'| \leq (1+2/c) |pq|$.  
\end{enumerate}
\end{lemma}
\begin{proof} 
The first claim follows from the following chain of inequalities: 
\begin{eqnarray*} 
  |pp'| & \leq & \diam(A) \\ 
  & \leq & \max ( \diam(A) , \diam(B) ) \\ 
  & \leq & (1/c) |AB| \\ 
  & \leq & (1/c) |pq|. 
\end{eqnarray*} 
The proof of the second claim is symmetric. For the proof of the 
third claim, we combine the first two claims and use the triangle 
inequality: 
\begin{eqnarray*} 
 |p'q'| & \leq & |p'p| + |pq| + |qq'| \\ 
      & \leq & (1/c) |pq| + |pq| + (1/c) |pq| \\ 
      & = & ( 1 + 2/c ) |pq| .
\end{eqnarray*}
\end{proof} 

\begin{definition}[Callahan and Kosaraju~\cite{ck-dmpsa-95}] 
Let $(S,|\cdot|)$ be a finite metric space and let $c>0$ be a real 
number. A \emph{well-separated pair decomposition (WSPD)} for $S$ is a 
sequence \[  \{ A_1,B_1 \} , \{ A_2,B_2 \} , \ldots, \{ A_m,B_m \}
\]
of pairs of non-empty subsets of $S$, for some integer $m$, such that 
\begin{enumerate}
\item for each $i$ with $1 \leq i \leq m$, $A_i$ and $B_i$ are 
well-separated, and
\item for any two distinct points $p$ and $q$ of $S$, there is exactly 
one index $i$ such that
\begin{enumerate}
\item $p \in A_i$ and $q \in B_i$, or
\item $p \in B_i$ and $q \in A_i$.
\end{enumerate}
\end{enumerate}
The integer $m$ is called the \emph{size} of the well-separated
pair decomposition.
\end{definition}

\subsection{The Faulty-Degree Spanner} 
\label{secFDS}

Let $(S,|\cdot|)$ be a finite metric space, let $f \geq 1$ be an integer, 
and let $\eps>0$ be a real number. Assume that
\[  \{ A_1,B_1 \} , \{ A_2,B_2 \} , \ldots, \{ A_m,B_m \}
\]
is a well-separated pair decomposition for $S$ with separation ratio 
\[ c = 2 + 4/\eps . 
\] 
We define a graph $G=(S,E)$ whose edge set $E$ is obtained as follows: 
For each $i$ with $1 \leq i \leq m$, define the following two subsets 
$A'_i$ and $B'_i$ of $A_i$ and $B_i$, respectively: 
\begin{itemize}
\item If $A_i$ has size at most $2f+1$, then $A'_i = A_i$. 
\item If $A_i$ has size at least $2f+2$, then $A'_i$ is an arbitrary 
subset of size $2f+1$ of $A_i$.  
\item If $B_i$ has size at most $2f+1$, then $B'_i = B_i$. 
\item If $B_i$ has size at least $2f+2$, then $B'_i$ is an arbitrary 
subset of size $2f+1$ of $B_i$.  
\end{itemize}
The edge set $E$ contains the edges of the complete bipartite graph 
with vertex sets $A'_i$ and $B'_i$. 

It is clear that the total number of edges in $E$ is at most 
$(2f+1)^2 m$. In the following subsection, we will prove that $G$ is 
an $f$-faulty-degree $(1+\eps)$-spanner for $S$. 

\subsection{Analysis} 
Let $F$ be an arbitrary subset of the edge set of $G$ such that the 
maximum degree of the graph $(S,F)$ is at most $f$. 
 
\begin{lemma} 
\label{lemWSPD}
For each edge $\{p,q\}$ in $K_S \setminus F$, we have 
\[ \delta_{G \setminus F}(p,q) \leq (1+\eps) \cdot |pq| . 
\]
\end{lemma} 
\begin{proof}
Let $i$ be the index in $\{1,2,\ldots,m\}$ such that (a) $p \in A_i$ 
and $q \in B_i$ or (b) $p \in B_i$ and $q \in A_i$. We may assume 
without loss of generality that (a) holds. 

The proof is by induction on the rank of the distance $|pq|$ in the set 
of all edge lengths in $K_S \setminus F$. 

For the base case, assume that $\{p,q\}$ is a shortest edge in 
$K_S \setminus F$. We first show that the size of the set $A_i$ is at 
most $2f$. Assume that this set has at least $2f+1$ points. Then $K_S$ 
contains at least $2f \geq f+1$ edges $\{p,p'\}$ for which $p' \in A_i$. 
At least one of these edges is in the graph $K_S \setminus F$. Since by 
Lemma~\ref{4insamepair}, $|pp'| < |pq|$, this is a contradiction. 
By a symmetric argument, we can show that the size of the set $B_i$ is 
at most $2f$. It follows that $p \in A_i = A'_i$ and 
$q \in B_i = B'_i$. Thus, $\{p,q\}$ is an edge in $G$. Since 
$\{p,q\} \not\in F$, it follows that 
\[ \delta_{G \setminus F}(p,q) = |pq| \leq (1+\eps) \cdot |pq| . 
\]

For the induction step, assume that $\{p,q\}$ is not a shortest edge in 
$K_S \setminus F$. If $\{p,q\}$ is in $G$, then it is in 
$G \setminus F$ and the lemma holds. Assume from now on that $\{p,q\}$ 
is not an edge in $G$. Then $p \not\in A'_i$ or $q \not\in B'_i$. 
We distinguish three cases. 

\vspace{0.5em} 

\noindent 
{\bf Case 1:} $p \not\in A'_i$ and $q \not\in B'_i$. 

We observe that each of $A_i$ and $B_i$ has at least $2f+2$ points, and  
each of $A'_i$ and $B'_i$ has exactly $2f+1$ points. 

Since $K_S$ contains $2f+1 \geq f+1$ edges $\{p,p'\}$ for which $p'$ is 
in $A'_i$, there is a point $p'$ in $A'_i$ such that $\{p,p'\}$ 
is in $K_S \setminus F$. Since, by Lemma~\ref{4insamepair}, 
$|pp'| \leq (1/c) |pq| < |pq|$, we have, by induction, 
\[ \delta_{G \setminus F}(p,p') \leq (1+\eps) \cdot |pp'| . 
\]
Note that, by the definition of $G$, for each point $q'$ in $B'_i$, 
the edge $\{p',q'\}$ is in $G$. Define 
\[ B''_i = \{ q' \in B'_i : \{p',q'\} \mbox{ is an edge in } 
             G \setminus F \} . 
\]
Since the size of $B'_i$ is $2f+1$, the size of $B''_i$ is at least 
$f+1$. Thus, $K_S$ contains at least $f+1$ edges $\{q,q'\}$ for which 
$q'$ is in $B''_i$. It follows that there is a point $q'$ in $B''_i$ such 
that $\{q,q'\}$ is in $K_S \setminus F$. Since, by Lemma~\ref{4insamepair}, 
$|qq'| \leq (1/c) |pq| < |pq|$, we have, by induction, 
\[ \delta_{G \setminus F}(q,q') \leq (1+\eps) \cdot |qq'| . 
\]
Note that, since $q'$ is in $B''_i$, $\{p',q'\}$ is an edge in 
$G \setminus F$. It follows that 
\begin{eqnarray*} 
 \delta_{G \setminus F}(p,q) & \leq & 
     \delta_{G \setminus F}(p,p') + |p'q'| + 
     \delta_{G \setminus F}(q',q) \\ 
 & \leq & (1+\eps) \cdot |pp'| + |p'q'| + (1+\eps) \cdot |q'q| . 
\end{eqnarray*} 
Using Lemma~\ref{4insamepair}, we conclude that 
\begin{eqnarray*} 
 \delta_{G \setminus F}(p,q) & \leq & 
    (1+\eps) \cdot (1/c) |pq| + (1+2/c) |pq| + 
      (1+\eps) \cdot (1/c) |pq| \\ 
  & = & ( 1 + (4+2\eps)/c ) |pq| \\
  & = & (1+\eps) |pq| ,
\end{eqnarray*} 
where the last equality follows from our choice for $c$. 

\vspace{0.5em} 

\noindent 
{\bf Case 2:} $p \in A'_i$ and $q \not\in B'_i$. 

In this case, $B_i$ has at least $2f+2$ points and $B'_i$ has exactly 
$2f+1$ points. By the definition of $G$, for each point $q'$ in $B'_i$, 
the edge $\{p,q'\}$ is in $G$. Define  
\[ B''_i = \{ q' \in B'_i : \{p,q'\} \mbox{ is an edge in } 
             G \setminus F \} . 
\]
Note that the size of $B''_i$ is at least $f+1$. Thus, $K_S$ contains 
at least $f+1$ edges $\{q,q'\}$ for which $q'$ is in $B''_i$. 
It follows that there is a point $q'$ in $B''_i$ such that $\{q,q'\}$ 
is in $K_S \setminus F$. By Lemma~\ref{4insamepair}, $|qq'| < |pq|$, 
implying that 
\[ \delta_{G \setminus F}(q',q) \leq (1+\eps) \cdot |q'q| . 
\]
Since $q'$ is in $B''_i$, $\{p,q'\}$ is an edge in $G \setminus F$. 
It follows that 
\[ \delta_{G \setminus F}(p,q) \leq 
    |pq'| + \delta_{G \setminus F}(q',q) \leq 
    |pq'| + (1+\eps) \cdot |q'q| .
\]  
As in Case~1, it follows that 
\[ \delta_{G \setminus F}(p,q) \leq (1+\eps) \cdot |pq| . 
\] 

\vspace{0.5em} 

\noindent 
{\bf Case 3:} $p \not\in A'_i$ and $q \in B'_i$. 

This case is symmetric to Case~2. 
\end{proof} 

We summarize the result of this section in the following theorem: 

\begin{theorem} 
\label{thmmainWSPD}  
Let $(S,| \cdot |)$ be a finite metric space, let $f \geq 1$ be an 
integer, and let $\eps>0$ be a real number. Assume there exists a 
well-separated pair decomposition for $S$ with separation ratio 
$c = 2 + 4/\eps$. Let $m$ be the size of this decomposition. 
Then the graph $G=(S,E)$ defined at the beginning of 
Section~\ref{secFDS} is an $f$-faulty-degree $(1+\eps)$-spanner for 
$S$ that has at most $(2f+1)^2 m$ edges. 
\end{theorem} 
\begin{proof} 
Let $F$ be an arbitrary subset of the edge set of $G$ such that the 
maximum degree of the graph $(S,F)$ is at most $f$. 
Let $p$ and $q$ be two distinct points in $S$. To prove that 
\[ \delta_{G \setminus F}(p,q) \leq (1+\eps) \cdot 
     \delta_{K_S \setminus F}(p,q) ,
\]
we consider a shortest path in $K_S \setminus F$ and apply 
Lemma~\ref{lemWSPD} to each edge on this path. 
\end{proof} 

We present some examples for which Theorem~\ref{thmmainWSPD} implies 
non-trivial results. Callahan and Kosaraju~\cite{ck-dmpsa-95} have 
shown that for any set $S$ of $n$ points in $\IR^d$, where the 
dimension $d$ is a constant and $|\cdot|$ is the Euclidean distance 
function, there is a WSPD for $S$ of size $m=O(n)$. Har-Peled and 
Mendel~\cite{hm-fcnld-05} generalized this this by showing that for 
every metric space of bounded doubling dimension, there is a WSPD for 
$S$ of size $m=O(n)$. Theorem~\ref{thmmainWSPD} implies that in both of 
these cases, we can construct an $f$-faulty-degree $(1+\eps)$-spanner 
with $O(f^2 n)$ edges. In the next section, we will improve the number of 
edges to $O(fn)$ for the Euclidean case.  

Let $S$ be a set of $n$ points in $\IR^d$, where $d \geq 2$ is a 
constant. The \emph{unit-disk graph} of $S$ is the graph with vertex set 
$S$, in which any two distinct points $p$ and $q$ are connected by an 
edge if their Euclidean distance $|pq|$ is at most one; the weight of 
this edge is equal to $|pq|$. The shortest-path distance in this graph 
defines a metric space. Note that the doubling dimension of this 
space is not bounded. Gao and Zhang~\cite{gz-wspdugm-03} have 
proved the following results: 

\begin{itemize}
\item For the case when $d=2$, there is a WSPD for the unit-disk graph 
metric on $S$ of size $m = O(n \log n)$. Theorem~\ref{thmmainWSPD} 
implies that there exists an $f$-faulty-degree $(1+\eps)$-spanner 
with $O(f^2 n \log n)$ edges. 
\item For the case when $d \geq 3$, there is a WSPD for the unit-disk 
graph metric on $S$ of size $m = O(n^{2-2/d})$. Theorem~\ref{thmmainWSPD} 
implies that there exists an $f$-faulty-degree $(1+\eps)$-spanner 
with $O(f^2 n^{2-2/d})$ edges. 
\end{itemize} 
 
\section{Euclidean Metric Spaces of Constant Dimension} 
\label{secEMS}

In this section, we consider the metric space $(S,| \cdot |)$, where 
$S$ is a set of $n$ points in $\IR^d$, for some constant $d$, 
and $| \cdot |$ is the Euclidean distance function.  

We will define two graphs with vertex set $S$ that are variants of the 
Yao-graph (see~\cite{fj-scdnng-81,y-cmstk-82}) and the $\Theta$-graph 
(see~\cite{c-aaspm-87,k-aceg-88}) and that depend on the maximum degree 
bound $f$ of any set $F$ of faulty edges.  
As we will see, for any real constant $\eps>0$, each of these graphs is 
an $f$-faulty-degree $(1+\eps)$-spanner with $O(fn)$ edges, where $n$ 
is the size of $S$. 

\subsection{Cones} 
\label{seccones} 
Let $\theta$ be a real number with $0 < \theta < \pi/4$ and let 
$\C$ be a collection of cones that cover $\IR^d$, such that 
for each cone $C$ in $\C$, the apex of $C$ is at the origin $0$ 
and the angular diameter of $C$ is at most $\theta$, i.e., 
\[ \max \left\{ \angle(0x,0y) : x,y \in C \setminus \{0\} \right\}  
            \leq \theta.
\]
Lukovszki~\cite{l-nrgst-99} has shown how to obtain such a collection 
$\C$ consisting of $O(1/\theta^{d-1})$ cones; 
see also~\cite[Chapter~5]{ns-gsn-07}.

\subsection{Yao- and Theta-Graphs} 
Let $S$ be a set of $n$ points in $\IR^d$ and let $f \geq 0$ be an 
integer. Let $\theta$ be a real constant with $0 < \theta < \pi/4$, and 
let $\C$ be a collection of cones as in Section~\ref{seccones}.   

For any cone $C$ in $\C$, we fix an arbitrary ray $\ell_C$ that 
emanates from the origin and is completely contained in $C$. 
For any point $p$ in $S$ and any cone $C$ in $\C$, let $C+p$ denote 
the cone with apex $p$ obtained by translating $C$, i.e.,
$C+p = \{ x+p : x \in C \}$. Similarly, we denote by $\ell_C + p$ 
the ray emanating from $p$ that is parallel to $\ell_C$. 
We denote by $S_{p,C}$ the set of all points in $S \setminus \{p\}$ 
that are in the cone $C+p$, i.e., 
$S_{p,C} = (C+p) \cap (S \setminus \{p\})$. 

\begin{figure}[t]
\centering
\includegraphics[scale=0.7]{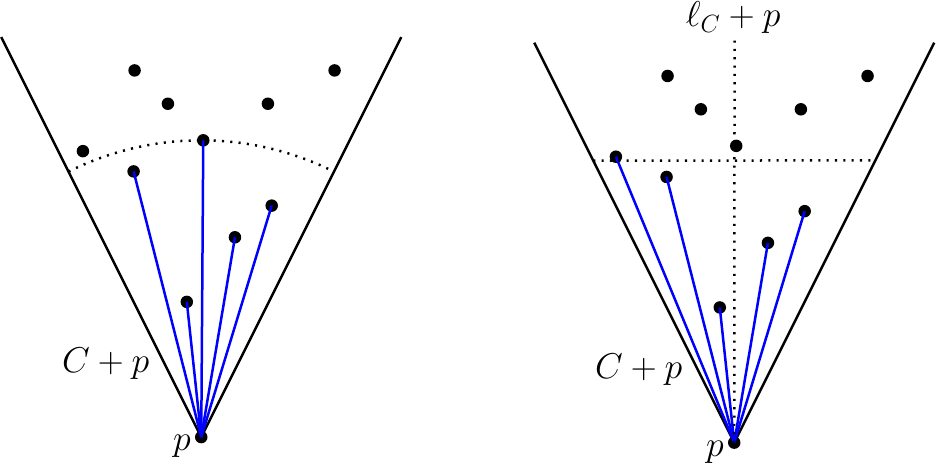}
\caption{The left figure illustrates Yao$(\theta,2f+1)$, where $f=2$. 
The graph has $2f+1=5$ (blue) edges in the cone $C+p$ that are incident 
on the point $p$; the endpoints of these edges are the five points in 
this cone that are closest to $p$. The right figure illustrates 
$\Theta(\theta,2f+1)$, where $f=2$ and $\ell_C + p$ is the bisector 
of the cone $C+p$. The graph has $2f+1=5$ (blue) edges 
in the cone $C+p$ that are incident on the point $p$; the endpoints of
these edges are the five points whose orthogonal projections onto the 
ray $\ell_C +p$ are closest to $p$.}
\label{figYT}
\end{figure}

We are now ready to define the variants of the Yao- and $\Theta$-graphs; 
refer to Figure~\ref{figYT}. 

\begin{enumerate} 
\item The graph Yao$(\theta,2f+1)$ has vertex set $S$. For each point 
$p$ in $S$ and for each cone $C$ in $\C$, we add an edge between $p$ and 
each of the $\min ( 2f+1 , |S_{p,C}| )$ points in $S_{p,C}$ that are
closest (with respect to the Euclidean distance) to $p$. 
\item The graph $\Theta(\theta,2f+1)$ also has vertex set $S$. For each 
point $p$ in $S$ and for each cone $C$ in $\C$, consider again the set 
$S_{p,C}$. We add an edge between $p$ and each of the 
$\min ( 2f+1 , |S_{p,C}| )$ points in $S_{p,C}$, whose orthogonal 
projections onto the ray $\ell_C + p$ are closest (with respect to 
the Euclidean distance) to $p$. 
\end{enumerate} 

It is clear that the number of edges in each of these graphs is at 
most $|\C| \cdot (2f+1) n = O(fn)$. 

\subsection{Analysis} 
In this subsection, we will prove that each of the graphs 
Yao$(\theta,2f+1)$ and $\Theta(\theta,2f+1)$ is an $f$-faulty-degree 
$t$-spanner for $S$, where $t$ only depends on the angle $\theta$. 

We remark that Lukovszki~\cite{l-nrftg-99} used $\Theta(\theta,f+1)$ 
for the case when at most $f$ edges can be removed. Our proof is 
similar. 

\begin{figure}[t]
\centering
\includegraphics[scale=0.7]{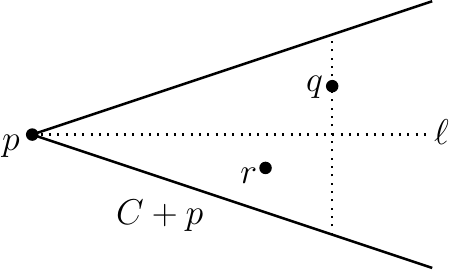}
\caption{Illustrating the first case in Lemma~\ref{lemmabasic}.} 
\label{figbasic}
\end{figure}

Our proof will use the following geometric result, which is illustrated 
in Figure~\ref{figbasic}.

\begin{lemma}   
\label{lemmabasic}
Let $p$ and $q$ be two distinct points in $S$, and let $C$ be a cone 
in $\C$ such that $q \in C+p$. 
\begin{enumerate} 
\item Let $\ell$ be an arbitrary ray that emanates from $p$ and is 
completely contained in $C+p$. Let $r$ be a point in $S_{p,C}$ such that 
the orthogonal projection of $r$ onto $\ell$ is at least as close to 
$p$ as the orthogonal projection of $q$ onto $\ell$. Then,
\[ |rq| \leq |pq| - (\cos\theta - \sin\theta) |pr| . 
\] 
\item Let $r$ be a point in $S_{p,C}$ such that $|pr| \leq |pq|$. Then,
\[ |rq| \leq |pq| - (\cos\theta - \sin\theta) |pr| . 
\] 
\end{enumerate} 
\end{lemma} 
\begin{proof}
The proof of the first claim can be found in~\cite[Lemma~4.1.4]{ns-gsn-07}. 
The second claim follows from the first claim, by taking for $\ell$ the 
ray that emanates from $p$ and goes through~$q$.  
\end{proof} 

In the rest of this subsection, we will prove that each of the graphs 
Yao$(\theta,2f+1)$ and $\Theta(\theta,2f+1)$ is an 
$f$-faulty-degree $t$-spanner for $S$, 
where $t = 1/(\cos\theta-\sin\theta)$. For both graphs, the proof will
be by induction. For Yao$(\theta,2f+1)$, the second claim in 
Lemma~\ref{lemmabasic} is used, whereas for $\Theta(\theta,2f+1)$, 
the first claim is used. Since both proofs are almost identical, we
only give the proof for $\Theta(\theta,2f+1)$. 

For simplicity, we denote the graph $\Theta(\theta,2f+1)$ by $\Theta$. 
Let $F$ be an arbitrary subset of the edge set of $\Theta$ and assume 
that the maximum degree of the graph $(S,F)$ is at most $f$. 

\begin{lemma} 
\label{lemma1} 
Let $\{p,q\}$ be an edge in $K_S \setminus F$ that is not in $\Theta$. 
Then, there is a point $r$ in $S_{p,C}$ such that 
\begin{itemize}
\item $r \neq q$, 
\item the orthogonal projection of $r$ onto the ray $\ell_C + p$ is at 
least as close to $p$ as the orthogonal projection of $q$ onto 
$\ell_C + p$, 
\item $\{p,r\}$ is an edge in $\Theta \setminus F$, and 
\item $\{r,q\}$ is an edge in $K_S \setminus F$. 
\end{itemize} 
\end{lemma} 
\begin{proof}
Since $\{p,q\}$ is not an edge in $\Theta$, this graph contains $2f+1$ 
edges $\{p,r\}$, where $r$ is in $S_{p,C}$ and $r \neq q$. For each such 
edge $\{p,r\}$, the orthogonal projection of $r$ onto the ray 
$\ell_C + p$ is at least as close to $p$ as the orthogonal projection 
of $q$ onto $\ell_C + p$. Since the degree of $p$ in $F$ is at most $f$, 
at least $f+1$ of these edges $\{p,r\}$ are in $\Theta \setminus F$. 
Observe that for each such edge $\{p,r\}$, the edge $\{r,q\}$ is in 
$K_S$. Since the degree of $q$ in $F$ is at most $f$, for at least one 
of these $f+1$ edges $\{p,r\}$ that are in $\Theta \setminus F$, the 
edge $\{r,q\}$ is in $K_S \setminus F$. 
\end{proof} 

\begin{lemma} 
\label{lemma2} 
For every edge $\{p,q\}$ in $K_s \setminus F$, 
$\delta_{\Theta \setminus F} (p,q) \leq t \cdot |pq|$, 
where $t = 1/(\cos\theta-\sin\theta)$.  
\end{lemma} 
\begin{proof}  
Let $C$ be a cone in $\C$ such that $q \in C+p$. We will prove the lemma 
by induction on the rank of the Euclidean distance $|pq|$ in the set of 
all edge lengths in $K_S \setminus F$. 

For the base case, assume that $\{p,q\}$ is a shortest edge in 
$K_s \setminus F$. We claim that this edge is in $\Theta$. If this 
is true, then it is in $\Theta \setminus F$ as well and 
$\delta_{\Theta \setminus F} (p,q) = |pq| \leq t \cdot |pq|$.  
Assume that $\{p,q\}$ is not in $\Theta$. Let $r$ be a point in $S_{p,C}$ 
as in Lemma~\ref{lemma1}. It follows from the first claim in 
Lemma~\ref{lemmabasic}, with $\ell = \ell_C + p$, that $|rq|<|pq|$, 
which contradicts the assumption that $\{p,q\}$ is a shortest edge in 
$K_s \setminus F$. 

From now on, assume that $\{p,q\}$ is not a shortest edge in 
$K_s \setminus F$. If this edge is in $\Theta$, then it is in 
$\Theta \setminus F$ and the lemma holds. Otherwise, let $r$ be a point 
in $S_{p,C}$ as in Lemma~\ref{lemma1}. As above, by 
Lemma~\ref{lemmabasic}, we have $|rq| < |pq|$. Thus, by induction, 
\[ \delta_{\Theta \setminus F} (r,q) \leq t \cdot |rq| . 
\]
It follows that 
\[ \delta_{\Theta \setminus F} (p,q) \leq 
    |pr| + \delta_{\Theta \setminus F} (r,q) \leq |pr| + t \cdot |rq| . 
\]
Using the first claim in Lemma~\ref{lemmabasic}, with $\ell = \ell_C + p$,
a straightforward calculation shows that 
\[ \delta_{\Theta \setminus F} (p,q) \leq t \cdot |pq| . 
\]
\end{proof} 

\begin{theorem} 
\label{YaoTheta} 
Let $S$ be a set of $n$ points in $\IR^d$, where $d$ is a constant, 
let $\eps > 0$ be a real constant, let $\theta$ be a constant with 
$0 < \theta < \pi/4$, such that $1/(\cos\theta-\sin\theta) \leq 1+\eps$, 
and let $f \geq 0$ be an integer. Each of the graphs Yao$(\theta,2f+1)$ 
and $\Theta(\theta,2f+1)$ is an $f$-faulty-degree 
$(1+\eps)$-spanner for $S$, having $O(fn)$ edges. 
\end{theorem} 
\begin{proof} 
We only give the proof for $\Theta = \Theta(\theta,2f+1)$. 
Set $t = 1/(\cos\theta-\sin\theta)$. 
Let $F$ be an arbitrary subset of the edge set of $\Theta$, such that
the maximum degree of the graph $(S,F)$ is at most $f$. 
Let $p$ and $q$ be two distinct points in $S$. If $\{p,q\}$ is an 
edge in $K_S \setminus F$, then, by Lemma~\ref{lemma2}, 
\[ \delta_{\Theta \setminus F} (p,q) \leq t \cdot |pq| = 
       t \cdot \delta_{K_s \setminus F}(p,q) \leq  
       (1+\eps) \cdot \delta_{K_s \setminus F}(p,q) . 
\]
If $\{p,q\}$ is not an edge in $K_S \setminus F$, then we consider a 
shortest path between $p$ and $q$ in $K_S \setminus F$, and apply 
Lemma~\ref{lemma2} to each edge on this path. 
\end{proof} 

\section*{Acknowledgement} 

We thank the anonymous referees for their careful reading of the first 
version of this paper, and for their detailed comments and suggestions.

\bibliographystyle{plain}
\bibliography{FaultDegreeTolerantSpanner}

\end{document}